\documentclass[pra,aps,twocolumn,floatfix,showpacs]{revtex4}
\usepackage{bm,graphicx,amsmath,bbm}
\usepackage{bbm}
\usepackage{times,txfonts,hyperref}
\usepackage{float}

\newcommand{\rp}[1]{(\ref{#1})}

\newcommand{\up}{\uparrow}
\newcommand{\dow}{\downarrow}
\newcommand{\br}[1]{\langle #1|}

\newcommand{\ke}[1]{|#1\rangle}

\newcommand{\pq}[1]{\left[ #1 \right]}
\newcommand{\pg}[1]{\left\{ #1 \right\}}

\newcommand{\lpt}[1]{\left( #1 \right.}

\newcommand{\rpt}[1]{\left. #1 \right)}

\newcommand{\Tr}{{\rm Tr}}

\newcommand{\nn}{{\nonumber}}

\begin{document}
 \title{Surface entanglement in quantum spin networks}
 \author{Stefano Zippilli, Salvatore Marco Giampaolo, Fabrizio Illuminati}
\affiliation{
Dipartimento di Ingegneria Industriale, Universit\`a degli Studi di Salerno,~Via Ponte don Melillo, I-84084 Fisciano (SA), Italy.
}

\date{February 5, 2013}

\begin{abstract}
We study the ground-state entanglement in systems of spins forming the boundary of a quantum spin network in arbitrary geometries and dimensionality. We show that as long as they are weakly coupled to the bulk of the network, the surface spins are strongly entangled, even when distant and non directly interacting, thereby generalizing the phenomenon of long-distance entanglement occurring in quantum spin chains. Depending on the structure of the couplings between surface and bulk spins, we discuss in detail how the patterns of surface entanglement can range from multi-pair bipartite to fully multipartite. In the context of quantum information and communication, these results find immediate application to the implementation of quantum routers, that is devices able to distribute quantum correlations on demand among multiple network nodes.
\end{abstract}


\maketitle

\section{Introduction}

Quantum spin networks are being actively investigated for quantum technological applications ranging from quantum teleportation to the quantum internet. Their dynamical properties have been extensively studied in order to assess their functionality for quantum state engineering and quantum state transfer~\cite{Christandl2004,Greentree,Yang,Friesen,Cubitt2008,Giampaolo1,Tufarelli,Kraus2008,Verstraete2009,Pemberton-Ross2011,Kay,
Karimipour,Ajoy,Nikolopoulos,Bayatnew}, while their static properties have been analyzed in order to identify ground or equilibrium thermal states useful as entangled quantum resources~\cite{Torma,Wang,Hutton,Rothlisberger2008,Facchi,Campos Venuti1,Campos Venuti2,GiampaoloLong1,GiampaoloLong2,Gualdi}. In principle, they can be simulated with a variety of atom-optical, molecular, and solid-state platforms~\cite{Buluta,Kim,Khromova,Candini,Zhang,Koppens2012,Yao2012}, which makes them extremely flexible models for the implementation of quantum technologies. On the other hand, a very important task in many fundamental protocols of quantum information and communication is the harnessing of entanglement between remote quantum objects. In particular, the distribution of entanglement between distant, not directly interacting systems, which may be part of a network, have been proposed and explored in many different physical settings~\cite{Kimble,Cirac,Bose2003,Eisert,kraus,Zippilli08,Brask,
Munro10,Gonzalez-Tudela,Leijnse,Banchi,Wolf,Trifunovic,Zippilli2013,Campos Venuti1,Campos Venuti2,GiampaoloLong1,GiampaoloLong2,Gualdi}. In this context, a relevant concept is that of long-distance entanglement (LDE)~\cite{Campos Venuti1,Campos Venuti2,GiampaoloLong1,GiampaoloLong2,
Gualdi}, namely the emergence of a sizeable end-to-end entanglement in spin chains whenever the coupling of the end spins to the remainder of the system is sufficiently weak. The occurrence of this intriguing phenomenon in one-dimensional systems naturally suggests to investigate the existence of analogous or more general effects in higher dimensions and nontrivial geometries.

In the present work we extend the notion of LDE~\cite{Campos Venuti1,Campos Venuti2,GiampaoloLong1,GiampaoloLong2,Gualdi} to higher dimensional lattices and generic geometries. We introduce the concept of {\em surface entanglement}, which refers to the emergence of entanglement in the reduced ground-state density matrix of the boundary spins of the network, that is the surface spins. In this respect, the LDE can be seen as a particular case of this more general phenomenon, corresponding to the case in which the surface coincides with two end spins of a liner spin chain. As we shall see, the main feature of surface entanglement, at odds with LDE, is that it can involve a large number of distant, non-interacting spins so that many different structures of the shared correlations can be identified between them, ranging from large collections of bipartite-entangled spin pairs to complex patterns of strong multipartite entanglement.
On the other hand, similarly to LDE, one of the conditions for the onset of surface entanglement is that the couplings between the spins of the surface and the ones in the bulk be much smaller than the coupling strengths among the spins in the bulk. This feature implies that surface
entanglement can be controlled by adjusting locally the interactions between the surface spins and the bulk, while the actual interactions within the rest of the network are up to a certain degree (as suitably specified below) irrelevant. This makes the present approach appealing for the realistic implementation of quantum routers~\cite{Pemberton-Ross2011,Plenio,Wojcik,Zueco,Chudzicki2010,Bayat}, namely devices able to distribute quantum correlations between distant nodes of a network of quantum systems.

The paper is organized as follows. In Sec.~\ref{Model} we introduce the class of quantum spin models to investigate, and we derive an effective Hamiltonian for the surface spin-dynamics in the limit of small coupling between surface and bulk spins, showing that in this limit the effective Hamiltonians share always the same symmetries of the original models. In Sec.~\ref{TwoSpins} we address the most elementary case in which the surface is made of only two spins, an immediate generalization of the LDE in linear chains to bulks of arbitrary dimensionality and geometry. In Sec.~\ref{MultiSpins} we consider the general situations in which the surface is made of an arbitrary number of spins, and we discuss how the tailoring of the interactions between surface and bulk spins can yield either large arrangements of bipartite entangled
spin pairs on the surface or complex patterns of multipartite entanglement involving many or even all of the surface spins. Finally, in Sec.~\ref{Conclusions} we draw our conclusions and discuss possible outlooks.

\section{Surface spins: total and effective Hamiltonians, and symmetries}\label{Model}

In this paper we analyze the properties of ground-state entanglement in various classes of spin-1/2 models for lattices of generic geometries and dimensions. We consider structures that can be characterized by two subsets of spins, respectively $B$ and $S$, where $B$ indicates the set of bulk spins and $S$ is the set of surface spins. These sets are not arbitrary; the set $B$ of bulk spins is assumed to be described by an Hamiltonian whose ground state is non-degenerate, thus implying that the total number of spins in the bulk is even; the set $S$ of surface spins is assumed to be such that the spins in $S$ do not interact among themselves, and each spin in $S$ interacts with a single spin in $B$ with a strength that is significantly smaller than the typical one between the spins in $B$ (see Figs.~\ref{fig1} and \ref{fig3} for examples). In order to be as general as possible we assume that the spins are coupled by anisotropic Heisenberg $XYZ$ exchange interactions. This ample class of models includes, as particular cases, the Ising, $XY$, $XX$, $XXZ$, and isotropic Heisenberg $XXX$ models.
Therefore, all the systems which we will discuss in the following are described by a total Hamiltonian of the following form
\begin{eqnarray}\label{H}
H_T = H_B + H_{SB},
\end{eqnarray}
where $H_B$ accounts for the interactions between the spins in the bulk, and $H_{SB}$ describes the interactions of the surface spins with the spins in the bulk. They are defined as
\begin{eqnarray}
H_B&=&\sum_{j,k\in B}\sum_{\alpha\in\pg{x,y,z}}
J_{jk}^\alpha\  \sigma_{B_j}^\alpha\sigma_{B_{k}}^\alpha \\
H_{SB}&=&\sum_{j\in S}\lambda_j\sum_{\alpha\in\pg{x,y,z}}
K_j^\alpha\  \sigma_{S_j}^\alpha\sigma_{B_{j}}^\alpha \, ,
\end{eqnarray}
where $\sigma_{S_j}^\alpha$ ($\sigma_{B_j}^\alpha$) with $\alpha\in\pg{x,y,z}$ are the standard
spin operators for the surface (bulk) spins. All the coupling strengths $J_{j,k}^\alpha$ and $K_j^\alpha$ are of the same order of magnitude. On the other hand, the bulk-surface interactions
are weighted by the coefficients $\lambda_j$ that can take any positive value in the range $[0,1]$.

An important preliminary result is that in the limit of $\lambda_j \ll 1$, the dynamics of the surface spins can be approximated by an effective interaction Hamiltonian between them that shares the same symmetries ($XY$, $XYZ$, $XXZ$, or $XXX$) of the original model Eq. (\ref{H}). In order to prove this property, let us consider a model for which all the weighting coefficients coincide:  $\lambda_j\equiv\lambda$
$\forall j\in S$, and let us define the set $\pg{\ke{\phi_\ell}}$ of eigenstates of $H_B$ and the corresponding eigenvalues $E_\ell$ (which satisfy $H_B\ke{\phi_\ell}=E_\ell\ke{\phi_\ell}$).
Under the assumptions that the ground state $\ke{\phi_0}$ of $H_B$ is non-degenerate and that
the gap between the ground and first excited state is much larger than $\lambda K_j^\alpha$ one can resort to perturbation theory in the parameter $\lambda$ in order to study the low-energy eigenstates of the total Hamiltonian $H_T = H_B + H_{SB}$. At zeroth order, the ground space of $H_T$ is degenerate with degeneracy equal to the dimension of the Hilbert space for the surface spins, and it is spanned by
the states of the form $\ke{\phi_0}\otimes\ke{\psi_p}$, where the states $\ke{\psi_p}$ form a basis in the Hilbert space of the surface spins. Thereby, the effective interaction Hamiltonian at second order in $\lambda$ for the surface spins reads
\begin{eqnarray}
H_{\rm eff}&=&-\sum_{\ell\neq 0} \frac{
\br{\phi_0}H_{SB}\ke{\phi_\ell}\br{\phi_\ell}H_{SB}\ke{\phi_0}}
{E_\ell-E_0}.
\end{eqnarray}
Taking into account that the anisotropic Heisenberg Hamiltonian $H_B$ in the bulk preserves the parity with respect to the three fundamental directions, i.e. $\pq{H_B,P_\alpha}=0$ with $P_\alpha=\prod_j \sigma_{B_j}^\alpha$ and $\alpha\in\pg{x,y,z}$,  we can rewrite $H_{\rm eff}$ as
\begin{eqnarray}
H_{\rm eff}&=&\sum_{\alpha\in\pg{x,y,z}}\sum_{j, k\in S} \Lambda_{j,k}^\alpha \sigma_{S_j}^\alpha\sigma_{S_k}^\alpha
\end{eqnarray}
where the effective coupling strength $\Lambda_{j,k}^\alpha$ is equal to
\begin{eqnarray}
\Lambda_{j,k}^\alpha&=&
-2\lambda^2 K_j^\alpha K_k^\alpha \sum_{\ell\neq 0} \frac { {\rm Re}
\br{\phi_{0}} \sigma_{B_j}^\alpha \ke{\phi_{\ell}}
\br{\phi_{\ell}} \sigma_{B_k}^\alpha \ke{\phi_{0}}
}{E_\ell-E_0}.
\end{eqnarray}
We see that this effective Hamiltonian describes, in general, a fully connected $XYZ$ model (or $XY$ model if $K_k^z \equiv 0$ $\forall k$) for the surface spins, that is each surface spin interacts with all other surface spins.

Moreover, if the bulk Hamiltonian $H_B$ commutes also with the total magnetization along the $z$-axes, $S_z=\sum_j\sigma_{B_j}^z$, for instance if it is a Hamiltonian of the
$XXZ$ form, then the coefficients $\Lambda_{j,k}^x$ and $\Lambda_{j,k}^y$ can be rewritten as
\begin{eqnarray}
\Lambda_{j,k}^x&=&
-2\lambda^2 K_j^x K_k^x \sum_{\ell\neq 0}
\frac { {\rm Re}
\br{\phi_{0}} \sigma_{B_j}^+ \ke{\phi_{\ell}}
\br{\phi_{\ell}} \sigma_{B_k}^- \ke{\phi_{0}}
}{E_\ell-E_0},
\nn\\
\Lambda_{j,k}^y&=&
-2\lambda^2 K_j^y K_k^y \sum_{\ell\neq 0}
\frac { {\rm Re}
\br{\phi_{0}} \sigma_{B_j}^+ \ke{\phi_{\ell}}
\br{\phi_{\ell}} \sigma_{B_k}^- \ke{\phi_{0}}
}{E_\ell-E_0},
\end{eqnarray}
where the $\sigma_{B_j}^+$ and $\sigma_{B_k}^-$ are the raising and lowering spin operators, and $\Lambda^y_{j,k}=\frac{K^y_jK^y_k}{K^x_jK^x_k}\Lambda^x_{j,k}$.
Thus, if also $H_{SB}$ is of the $XXZ$ type, with $K^y_k\equiv K^x_k \, \forall k$, then also the effective Hamiltonian $H_{\rm eff}$ is of the $XXZ$ type.
Finally, in the case that the original total Hamiltonian $H_T$ belongs to the class of isotropic Heisenberg $XXX$ models, which commute with the total magnetization with respect to any axes, applying the same reasoning as in the previous cases, it is fairly straightforward to show that also in this instance the effective Hamiltonian is of the $XXX$ type.

In conclusion, the effective interaction Hamiltonian $H_{\rm eff}$ for the surface spins shares the same symmetries of the original total Hamiltonian $H_T$, and it is in general fully connected, regardless of the range of the interactions in the original model. As a consequence, on the one hand, the ground-state properties of  $H_{\rm eff}$ will be dictated by the symmetries of $H_T$ and, on the other hand, given the fully connected nature of $H_{\rm eff}$, one can anticipate that in general its ground state will exhibit complex patterns of bipartite and multipartite entanglement. These qualitative picture is confirmed quantitatively by the results of an extended numerical analysis performed directly on the original model $H_T$ and reported in Section \ref{NumRes}.

\section{Numerical results: exact diagonalization of the total Hamiltonian}
\label{NumRes}

In this section we analyze the entanglement properties of the reduced ground-state density matrix of the surface spins, obtained by tracing out the degrees of freedom of the bulk spins after exact numerical diagonalization of the total Hamiltonian $H_T$ for different dimensions, geometries, and symmetries.
We will also analyze in detail the energy gap between ground and first excited state, because this quantity is of central importance in assessing the thermal stability of the ground-state entanglement, in particular of that of the surface spins. Furthermore, in the perspective of an experimental realization of surface entanglement based on adiabatic ground-state preparation protocols~\cite{Farhi}, the size of the gap sets a limit to the maximum allowed velocity for adiabatic manipulation.

\subsection{Two-spin surface}\label{TwoSpins}

\begin{figure}[!t]
\begin{center}
\includegraphics[width=7cm]{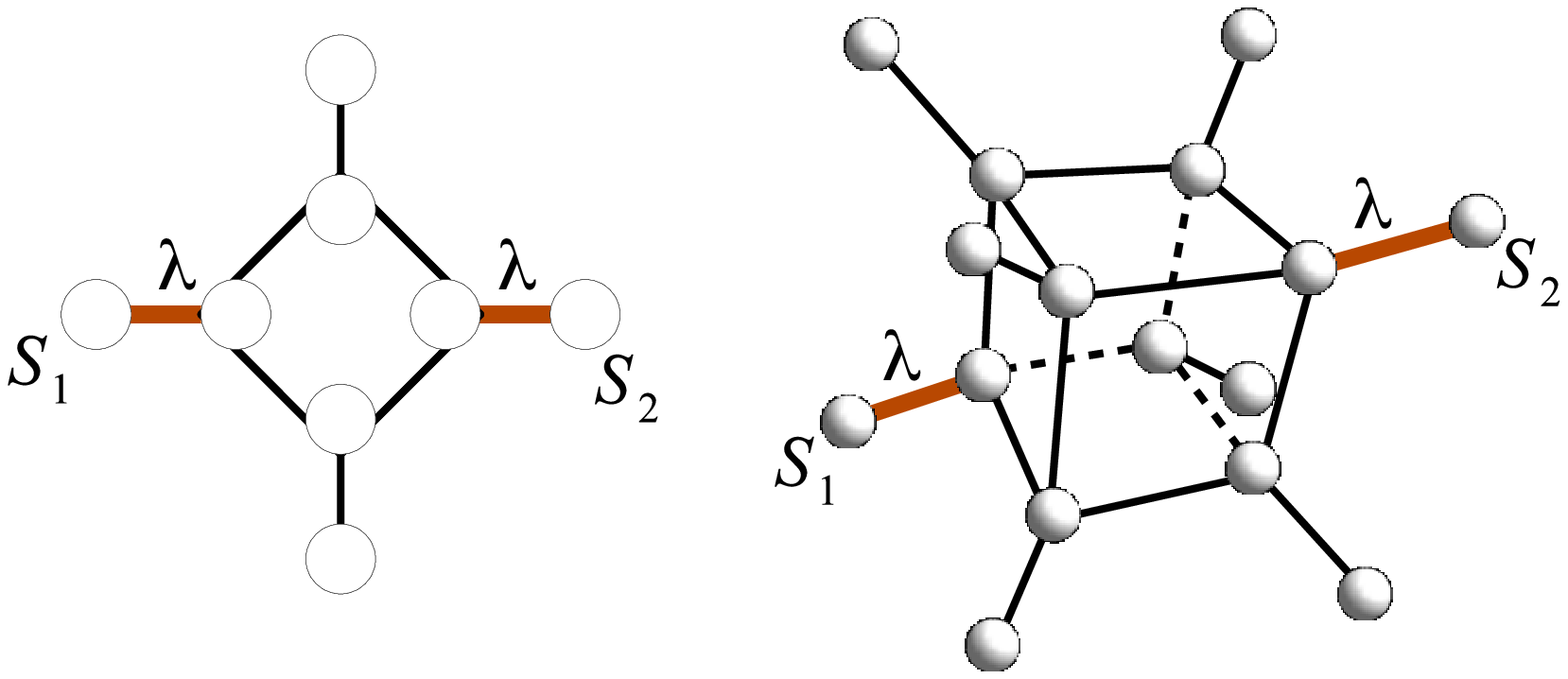}
\includegraphics[width=8cm]{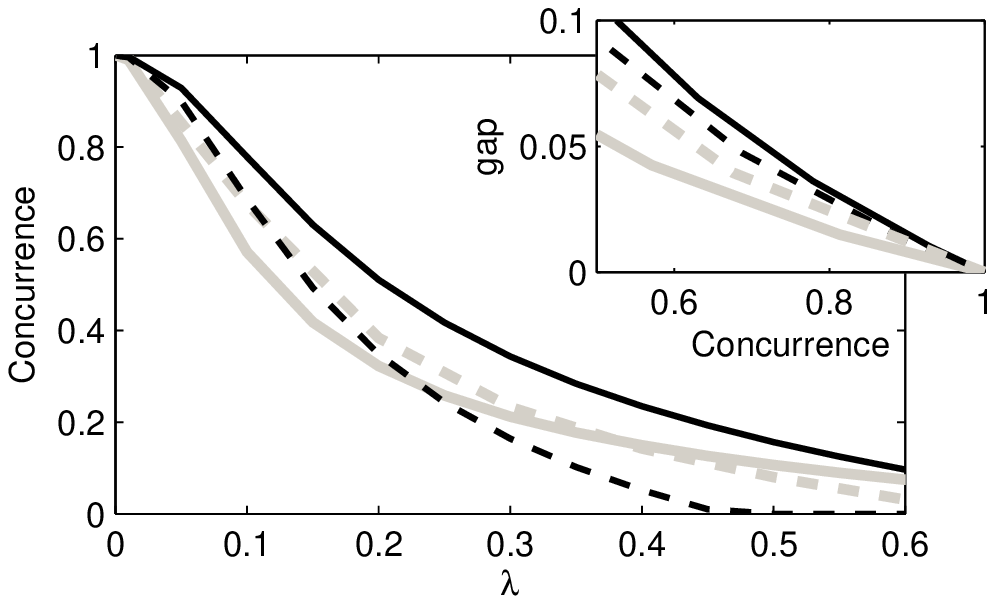}
\end{center}
\caption{Top: two possible simple schemes of spin networks with a surface consisting of only two spins. Left: a two-dimensional square configuration. Right: a three-dimensional cubic configuration. Spins are denoted by circles, while the lines connecting the circles indicate the spins that interact directly. The bulk spins interact with a rescaled dimensionless coupling normalized to unity. The couplings $\lambda$ identify the two surface spins $\pg{S_1,S_2}$, i.e. the (not directly interacting) spins at the boundary of the network that interact weakly with the bulk. Bottom: Concurrence between the surface-spins $S_1$ and $S_2$ as a function of $\lambda$ for the square (black lines) and the cubic (gray lines) configuration. These results correspond to $XX$ interactions (solid lines) with  $J_{jk}^x=J_{jk}^y=K^x_\ell=K^y_\ell=J$, $\forall j,k\in B$ and
$\forall \ell\in S$, and $XXZ$ interactions (dashed lines), with $2J^z_{jk}=2K^z_{\ell}=J_{jk}^x=J^y_{jk}=K^x_\ell=K^y_\ell=J$,
$\forall j,k\in B$ and $\forall \ell\in S$. In both cases the interactions are antiferromagnetic ($J>0$).  Inset: energy gap as a function of the two-spin surface concurrence.
}
\label{fig1}
\end{figure}
We consider first the simplest situation in which the surface is made of only two spins. This situation can be seen as a generalization of the LDE in quantum spin chains~\cite{Campos Venuti1,Campos Venuti2,GiampaoloLong1,GiampaoloLong2} to networks whose bulk has an arbitrary geometry but a minimal
number ($j=1,2$) of surface spins, defined as the - non directly interacting - boundary spins that interact weakly with the bulk via couplings $\lambda_j$. In the limit of $\lambda_j \ll 1$, and in agreement with the findings of the previous section, the two-spin ground-state reduced density matrix approaches a maximally entangled Bell state, with the exception of the cases in which the ground state of the two-spin effective Hamiltonian is degenerate, as in the Ising and the ferromagnetic isotropic Heisenberg models. Examples of results for the two-spin surface entanglement are reported, for two- and three-dimensional systems, and for $XX$ and $XXZ$ interactions, in Fig.~\ref{fig1}. We observe that in all cases, when
$\lambda_1=\lambda_2\equiv\lambda$ is sufficiently small, the concurrence of the reduced density matrix of the two surface spins approaches unity, meaning that in this limit the reduced state correctly approaches a Bell state as was to be expected from the discussion in the previous section. Moreover, since the effective coupling strengths $\Lambda_{1,2}^\alpha$ are proportional to $\lambda^2$, the energy gap tends to vanish as $\lambda$ is reduced, thus making the surface entanglement extremely unstable against thermal fluctuations in the limit of vanishing $\lambda$. On the other hand, as the concurrence decreases for larger $\lambda$, the energy gap increases correspondingly.

\begin{figure}[!t]
\begin{center}
\includegraphics[width=7cm]{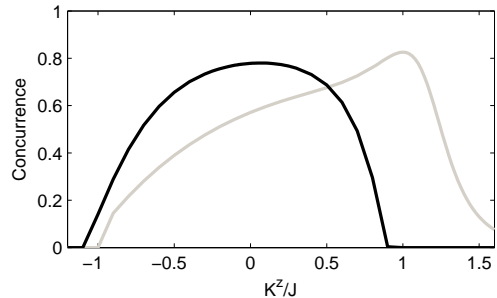}
\end{center}
\caption{Concurrence in the reduced ground state of the surface spins $S_1$ and $S_2$ for square (black line) and cubic (gray line) network configurations of Fig.~\ref{fig1} with $XXZ$ interactions ($J_{jk}^x=J_{jk}^y=K^x_\ell=K^y_\ell=J>0$, $\forall j,k\in B$ and
$\forall \ell\in S$), as a function of $K^z\equiv K^z_{\ell}$, $\forall \ell\in S$. Both curves are plotted for $\lambda=0.1$.}
\label{fig2}
\end{figure}

One notices from Fig. \ref{fig1} that, depending on the geometry, an anisotropy in the spin-spin interaction along the transverse $z$ direction may be instrumental to the surface entanglement. In particular, the cubic network configuration with $XXZ$ interactions possesses larger entanglement with respect to the same geometry with $XX$ interactions. In general, the optimal value of $K^z$ which maximize the entanglement is geometry-dependent. For example, as shown in Fig.~\ref{fig2}, at fixed $\lambda$
and in the case of the cubic configuration of Fig.~\ref{fig1}, the concurrence reaches its maximum, at $K^z\simeq J$, i.e. for $XXX$ isotropic Heisenberg interactions, whereas for the square configuration it is maximal at $K_z\sim 0.065\, J$.

\begin{figure}[!t]
\begin{center}
\includegraphics[width=8cm]{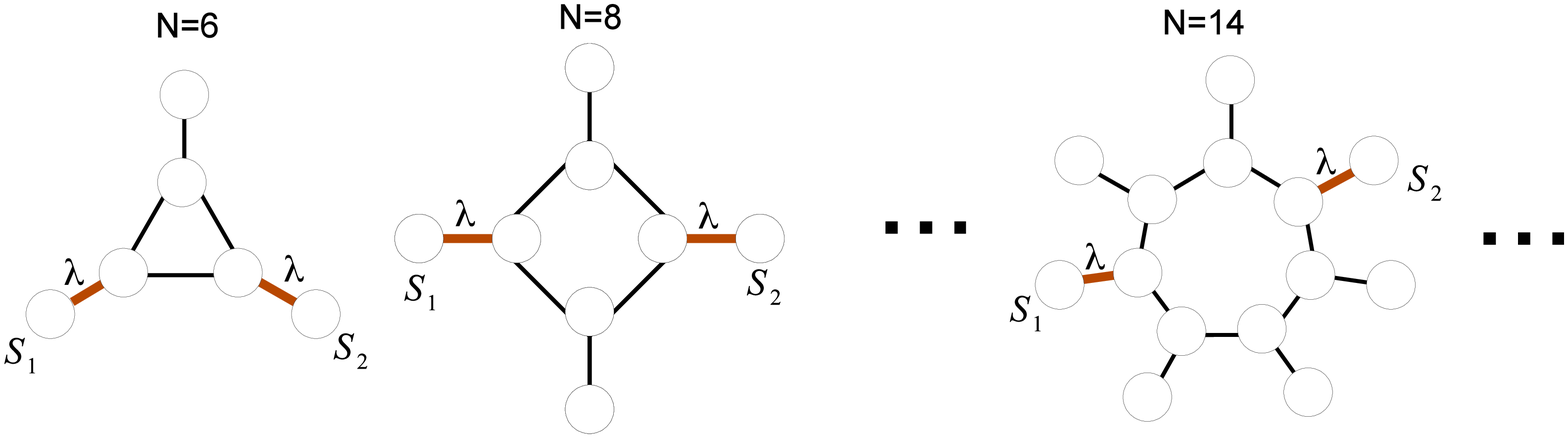}
\includegraphics[width=8cm]{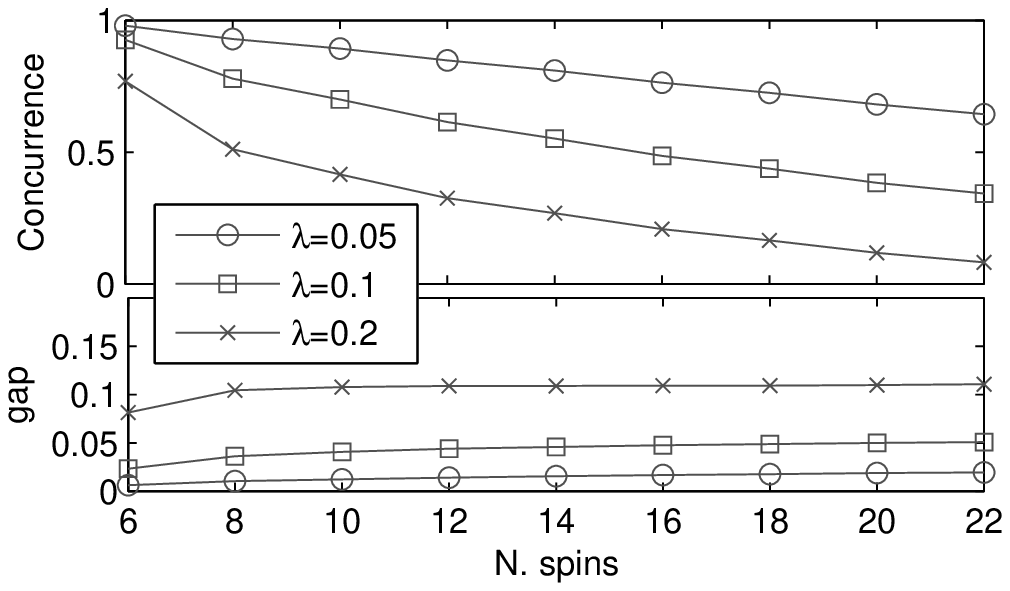}
\end{center}
\caption{Top: ring-shaped quantum spin network of increasing size. Bottom: surface concurrence (upper panel) and energy gap (lower panel) as functions of the total number of spins for this type of network. The parameter $\lambda$ denotes the weak coupling between the two surface spins and the bulk of the network. The interactions are of $XX$ ferromagnetic type with $J_{jk}^x=J_{jk}^y=K^x_\ell=K^y_\ell=J$, $\forall j,k\in B$ and $\forall \ell\in S$.}
\label{fig3}
\end{figure}

For assigned geometry, symmetries of the interactions, and surface-to-bulk coupling $\lambda$, the concurrence slowly decays with the dimension of the bulk (i.e. the number of spins in the lattice, excluding the two surface spins). This is shown in Fig.~\ref{fig3}, where the surface concurrence is reported for diverse models whose bulks have a ring-like shape with increasing number of spins. Maximum entanglement is obtained for the smallest (in terms of number of spins) configurations. Viceversa, the gap mildly increases with the number of spins. Therefore, this geometry results promising for an actual implementation of a rooter based on surface entanglement: depending on the desired working point, it entails a constrained optimization of the two contrasting requests of significantly sizeable entanglement and energy gap.

\begin{figure}[!t]
\begin{center}
\includegraphics[width=6cm]{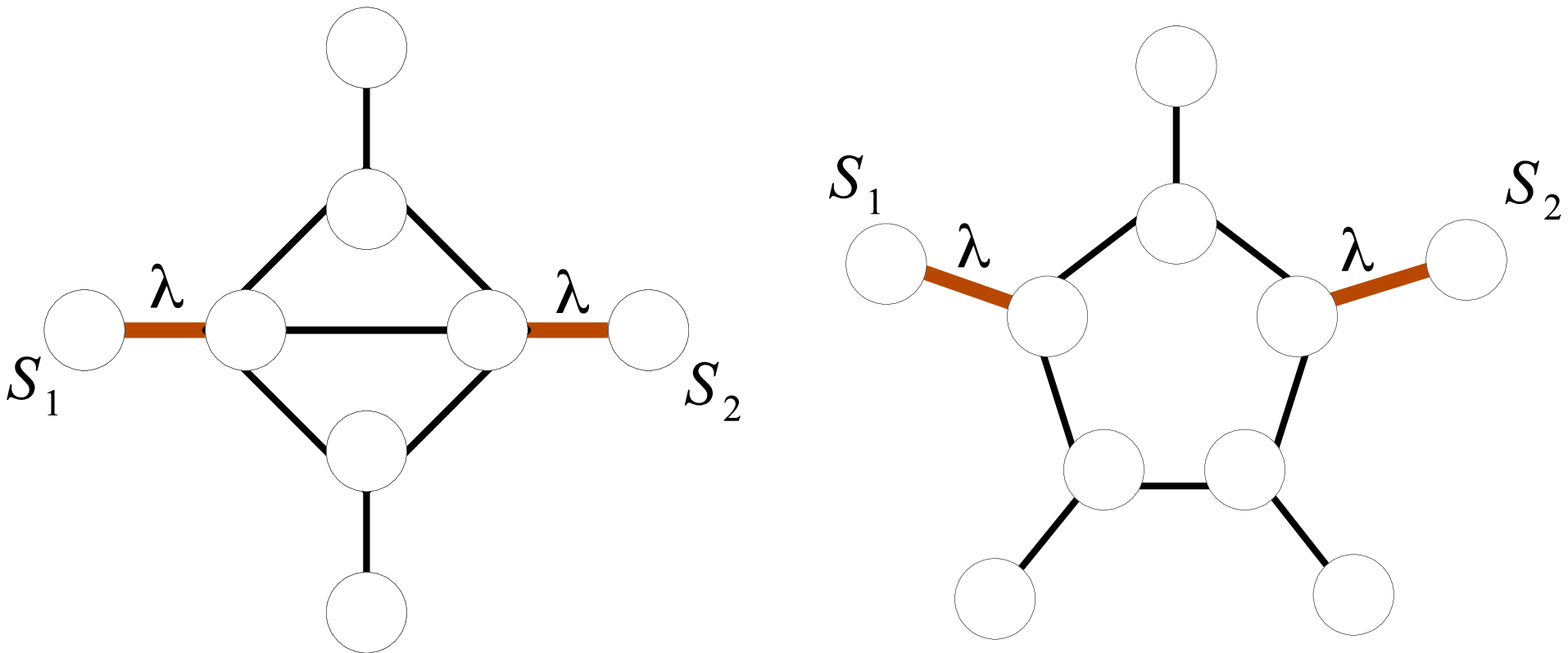}
\includegraphics[width=8cm]{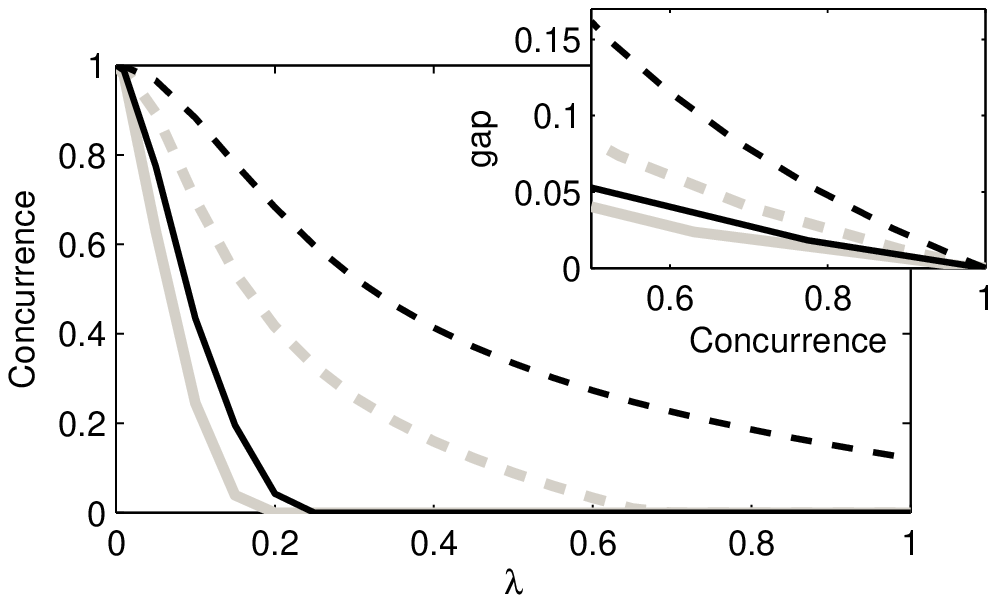}
\end{center}
\caption{Top: examples of two-dimensional networks with bulk configurations allowing for geometrical frustration. Left: square with five interaction bonds. Right: pentagonal with five interaction bonds. Bottom: concurrence between the surface spins $S_1$ and $S_2$ as a function of the surface-to-bulk coupling $\lambda$. The interactions are of the $XX$ type ($J_{jk}^x=J_{jk}^y=K^x_\ell=K^y_\ell=J$, $\forall j,k\in B$ and $\forall \ell\in S$). Solid lines: antiferromagnetic, geometrically frustrated case. Dashed lines: ferromagnetic, geometrically unfrustrated case. Black lines: network with square geometry of the bulk. Gray lines: network with pentagonal geometry of the bulk. Inset: corresponding behavior of the energy gap as a function of $\lambda$.}
\label{fig4}
\end{figure}

It should be noticed that the models studied in Figs.~\ref{fig1} and \ref{fig3} satisfy the generalized Toulouse criteria for frustration-free systems, as introduced in~\cite{Giampaolo2011}, and therefore they
are not geometrically frustrated. In general, a model that does not satisfy the generalized Toulouse criteria, and that therefore is geometrically frustrated, has a surface-spin concurrence which decays much faster with increasing surface-to-bulk coupling $\lambda$, as compared to the corresponding non frustrated model. This behavior is illustrated in Fig.~\ref{fig4} where models that do not satisfy the generalized Toulouse criteria are realized with anti-ferromagnetic nearest-neighbor interactions on square and pentagonal bulk geometries with an odd number (five) of interaction bonds. These are compared with the corresponding ferromagnetic models, which satisfy the generalized Toulouse criteria and are thus geometrically unfrustrated~\cite{Giampaolo2011}. Therefore, quantum spin networks with geometrically non frustrated bulk configurations should always be preferred for an optimization of the corresponding surface entanglement.

\begin{figure}[!t]
\begin{center}
\includegraphics[width=8cm]{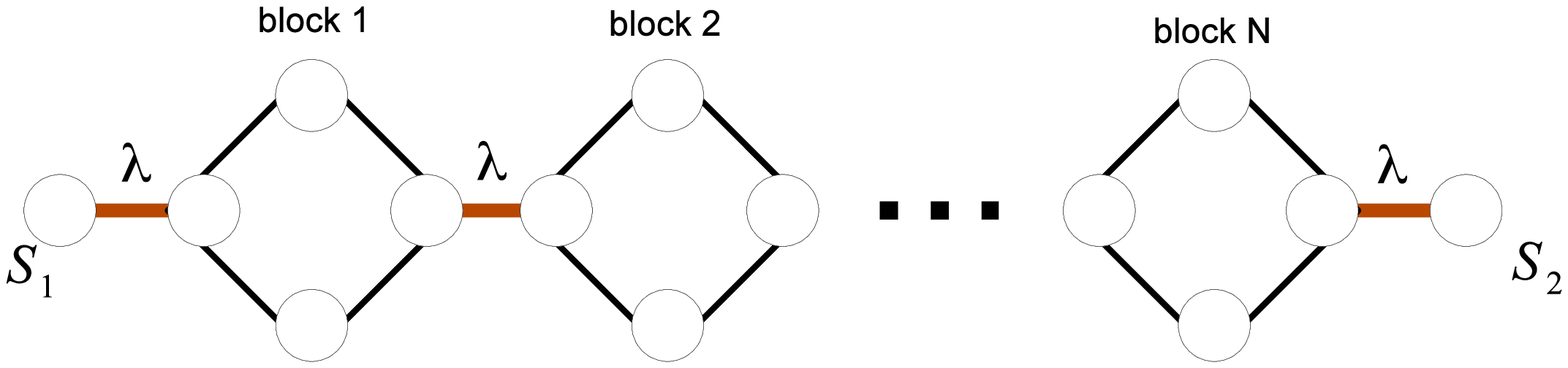}
\includegraphics[width=8cm]{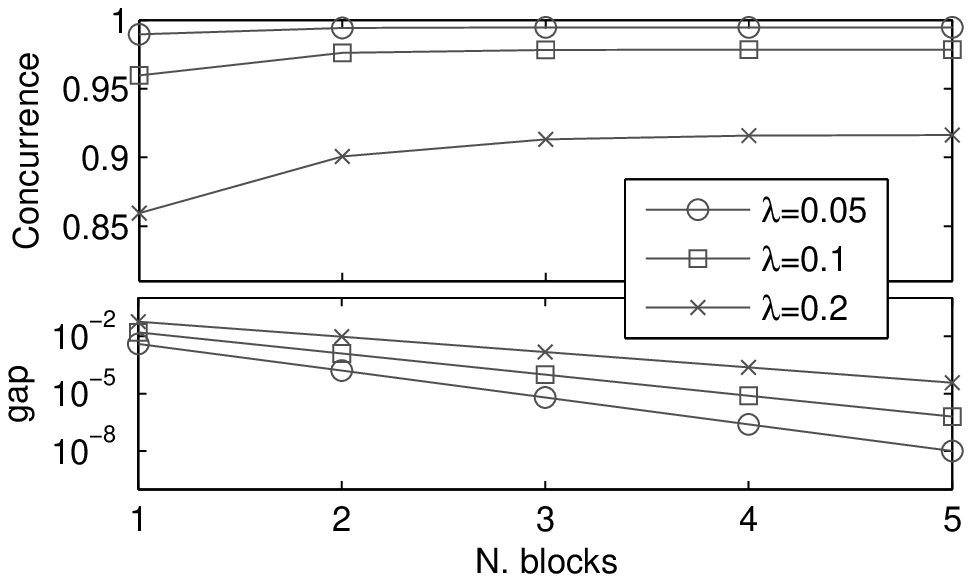}
\end{center}
\caption{Top: Quantum spin network with a modular structure of coupled two-dimensional blocks of spins. Bottom: for this type of modular network, ground-state concurrence between the surface spins $S_1$ and $S_2$ (upper panel) and corresponding energy gap (lower panel) as functions of the total number of blocks for different values of the weak coupling $\lambda$. The interactions are of the $XX$ type with $J_{jk}^x=J_{jk}^y=K^x_\ell=K^y_\ell=J$, $\forall j,k\in B$ and $\forall \ell\in S$, and the same results hold for ferromagnetic and antiferromagnetic interactions.
}\label{fig5}
\end{figure}

Finally, before considering the general case of networks with an arbitrary number of boundary spins, we turn to the concept of modular entanglement recently introduced for spin chains~\cite{Gualdi}, and we look for generalizations to higher dimensions and generic geometries. It has been demonstrated in Ref.~\cite{Gualdi}  that, in the case of linear spin chains, the properties of the end-to-end entanglement, including its stability against thermal fluctuations, can be greatly enhanced when various sub-chains are serially coupled by weak interactions to form a modular structure. Moving to spin networks with only two surface spins, a similar behavior is observed in Fig.~\ref{fig5}, where we consider an example in which the bulk of the network is organized in linearly coupled two-dimensional modules (blocks). Indeed, the associated surface entanglement is enhanced as the number of modules is increased, while the energy gap correspondingly decreases. The mechanism of modular entanglement can thus be effectively generalized to higher-dimensional geometries. Comparing Fig.~\ref{fig5} and Fig.~\ref{fig3} one sees that ring-like and modular networks exhibit opposite behaviors of surface entanglement and energy gap as a function of the number of bulk spins (bulk modules). This observation suggests the alternative use or the combination of the two configurations depending on the assigned physical or communication task.

\subsection{Many-spin surfaces}\label{MultiSpins}

So far we have considered networks whose surfaces consist of only two spins. When the boundary contains more than two spins, then many different patterns of ground-state surface entanglement can be identified. In particular, in the following we will consider two relevant limiting cases. Firstly we shall study the situation in which the surface spins dimerize in a series of strongly entangled pairs (bipartite entanglement replicated in a large number of entangled pairs). Secondly, we address the case in which the boundary is characterized by strong multipartite entanglement involving essentially all of the surface spins.
These two limiting cases are important because most of the intermediate possibilities, where complex patterns of bipartite and multipartite entanglement may coexist simultaneously, exhibit essentially various aspects of these two instances. In order to clarify this complex phenomenology we discuss separately the two limiting cases.

\subsubsection{Many-pair bipartite surface entanglement}

\begin{figure}[!t]
\begin{center}
\includegraphics[width=6.5cm]{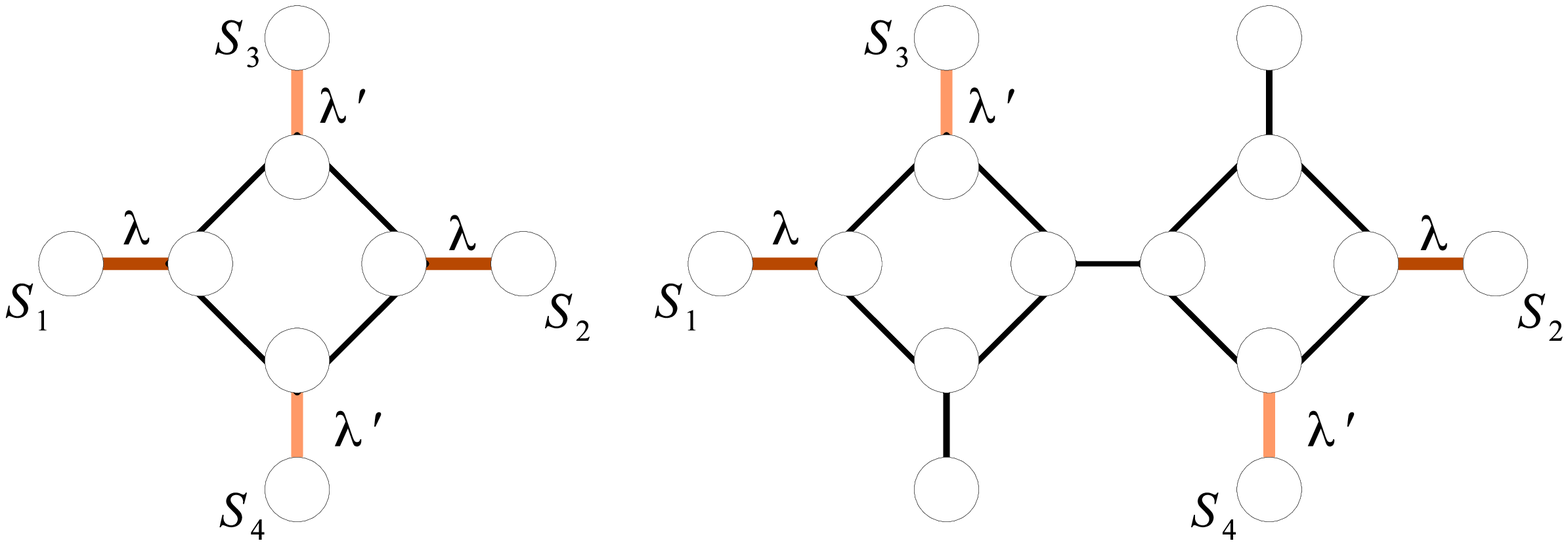}
\includegraphics[width=8cm]{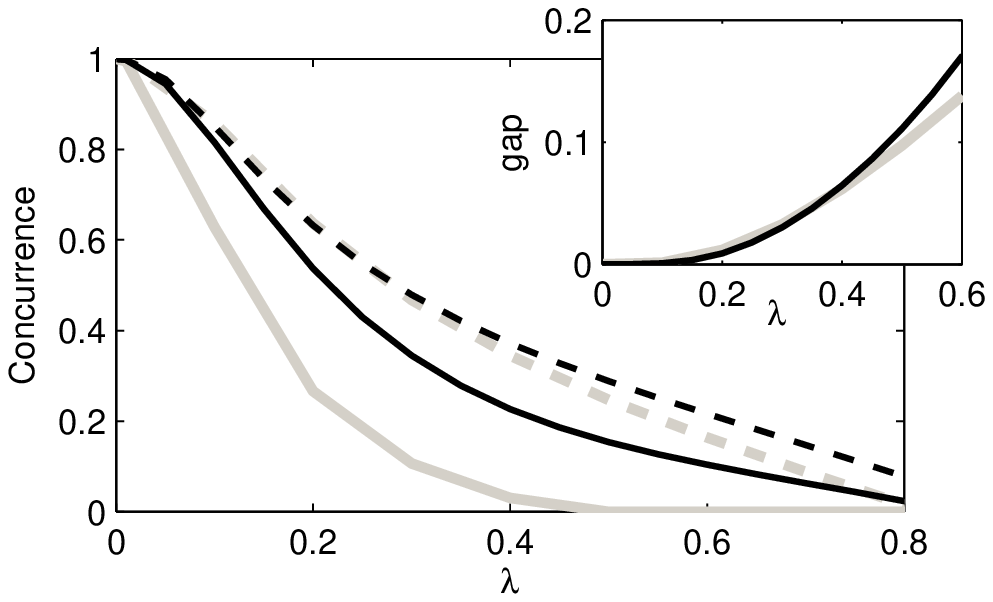}
\end{center}
\caption{Top: single-square (left) and nested tow-squares (right) two-dimensional architectures for quantum spin networks. Bottom: concurrence in the reduced ground state of the pairs $S_1-S_2$ (solid lines) and $S_3-S_4$ (dashed lines) of surface spins for the single-square (black lines) and the nested two-squares (gray lines) configurations. The interactions are of the $XX$ type with $J_{jk}^x=J_{jk}^y=K^x_\ell=K^y_\ell=J$, $\forall j,k\in B$ and $\forall \ell\in S$, with the same results holding for ferromagnetic and antiferromagnetic couplings. The surface spins $S_1$ and $S_2$ are coupled to the bulk with strength $\lambda$, while the surface spins $S_3$ and $S_4$ with strength $\lambda'=\lambda^2$. Inset: behavior of the energy gap as a function of $\lambda$.}
\label{fig6}
\end{figure}

\begin{figure}[!t]
\begin{center}
\includegraphics[width=4cm]{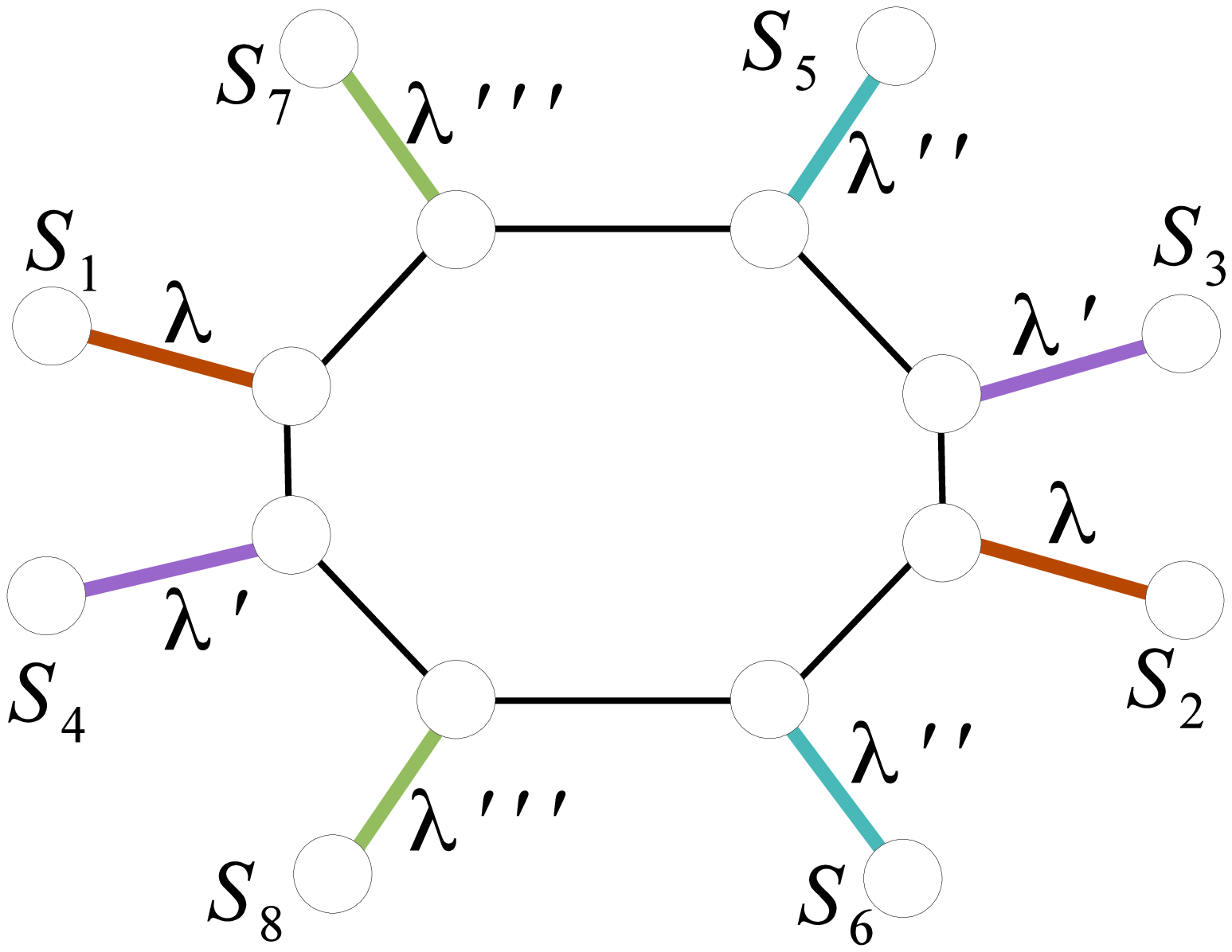}
\includegraphics[width=8cm]{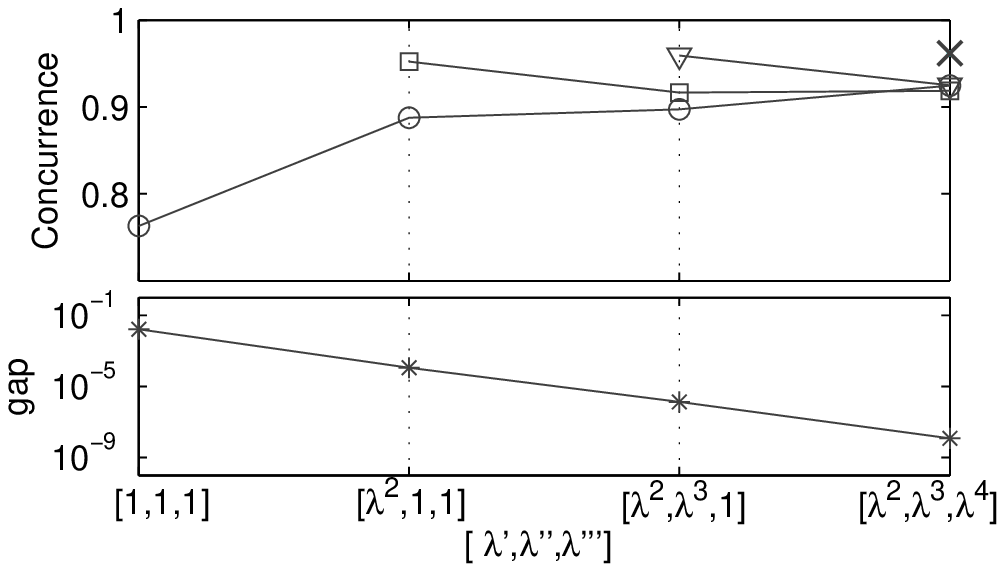}
\end{center}
\caption{
Top: Ring-shaped network. Bottom (upper panel): for this type of network, the concurrences in the reduced ground states of the different surface spin pairs: $S_1-S_2$ (circles), $S_3-S_4$ (squares), $S_5-S_6$ (triangles) and $S_7-S_8$ (crosses), as functions of different values of the couplings
$\lambda'$,$\lambda''$ and $\lambda''' $ as indicated in the abscissa, with the coupling $\lambda$
fixed at the value $\lambda=0.05$. The interactions are of the $XX$ type with $J_{jk}^x=J_{jk}^y=K^x_\ell=K^y_\ell=J$, $\forall j,k\in B$ and $\forall \ell\in S$, and the
same results hold for ferromagnetic and antiferromagnetic couplings. Bottom (lower panel): the
corresponding energy gap.
}
\label{fig7}
\end{figure}

Many entangled spin pairs are realized in the reduced ground state of the surface spins when their interaction with the bulk follows a two by two pattern such that every two surface spins share the same coupling $\lambda$ which in turn differs from that of all other pairs, as illustrated in Figs.~\ref{fig6} and \ref{fig7}. This result can be understood as an extension of the situation holding for two-spin surfaces - discussed in Sec.~\ref{TwoSpins} - to a series of many nested two-spin surfaces where each added term is coupled to the bulk with decreasing coupling strength. Let us, for example, consider the case of four surface spins, such that two of them are coupled to the bulk with a strength $\lambda$, whereas the coupling strength of the other two is $\lambda'$, and they satisfy the relation $\lambda' \ll \lambda \ll 1$. In complete agreement with what we have proved in Sec.~\ref{TwoSpins}, the two spins with the weakest interactions $\lambda'$ become strongly entangled. Indeed, they can be thought as forming the surface of an enlarged bulk consisting of the actual bulk and the two remaining spins in the surface. In turn, also the enlarged bulk can be seen as a reduced system with two surface spins that interact with the bulk with strength $\lambda$, and thus also the reduced ground state of this pair is strongly entangled.
In conclusion, the total surface of the network separates (dimerises) in two entangled spin pairs that become maximally entangled in the limit of vanishing surface-to-bulk couplings. This result extends straightforwardly to surfaces with many spins, resulting in a structure of many entangled dimers with the inclusion of an arbitrary number of spin pairs in the surface of the network.

%

We have verified this picture thoroughly by exact numerical diagonalization and the results are summarized in Figs.~\ref{fig6} and \ref{fig7}. The results for the basic configuration with four surface spins are shown in Fig.~\ref{fig6} for interactions of the $XX$ type. The concurrence of the surface spin pairs is
reported as a function of the two weak couplings $\lambda$ for the first pair and $\lambda' < \lambda$
for the second pair. Both pairs approach a maximally entangled Bell state in the limit of a vanishing coupling strength, thus demonstrating the potentiality of the surface entanglement as a resource for the parallel distribution of quantum correlations among many nodes of spin networks. The multi-pair bipartite entanglement occurs also when several pairs are considered, as illustrated in Fig.~\ref{fig7}, where we consider up to four entangled pairs in the reduced ground state of the network's surface, and we report the bipartite entanglement of four different pairs of surface spins with couplings in decreasing order $\lambda \geq \lambda' \geq \lambda'' \geq \lambda'''$. We see that as the values of the couplings are progressively reduced in hierarchical order, all the two-spin concurrences approach unity and tend to realize a collection of perfect Bell singlets. However, in the multi-pair case the energy gap decays rapidly as the number of pairs increases, thus making this configuration extremely sensitive to thermal fluctuations.

\subsubsection{Multipartite surface entanglement}

We have seen that a structure of many entangled pairs, approaching a dimerized configuration of many Bell singlets occurs when the coupling parameters $\lambda_j$ between the different surface spins and the bulk of the network are significantly different from each other. Let us now consider the opposite situation, in which essentially all the surface spins interact with the bulk spins with the same weak coupling: $\lambda_j\equiv\lambda\ll1$, $\forall j\in S$.

%
%
In this case, as discussed in Sec.~\ref{Model}, the effective Hamiltonian for the surface spins dynamics is that of a fully connected system, in the class of the Lipkin-Meshkov-Glick model~\cite{Lipkin}. The symmetries of the geometry of the effective model imply that the ground-state correlations have to be shared between all the spins, hence resulting in a long-range pattern of multipartite entanglement.
%
%
%
\begin{figure}[!t]
\begin{center}
\includegraphics[width=6cm]{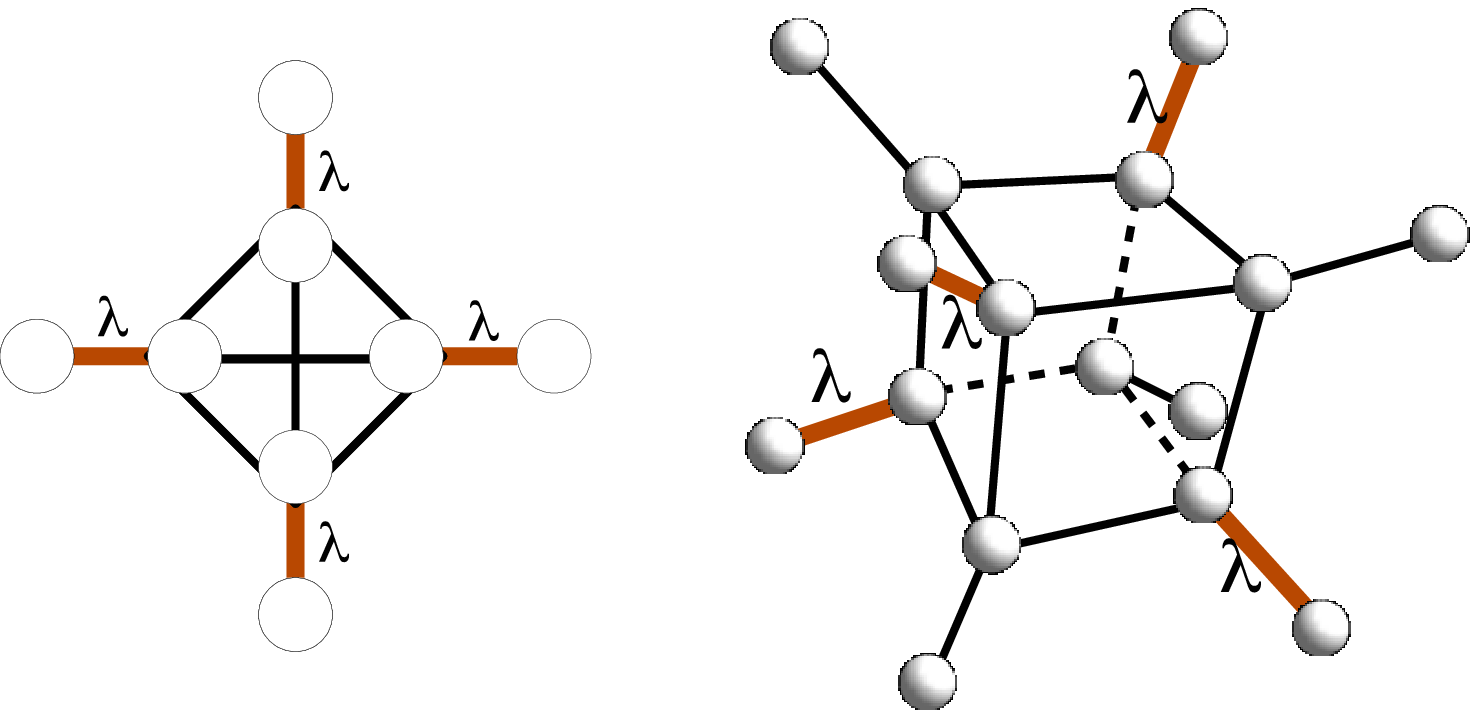}
\includegraphics[width=8cm]{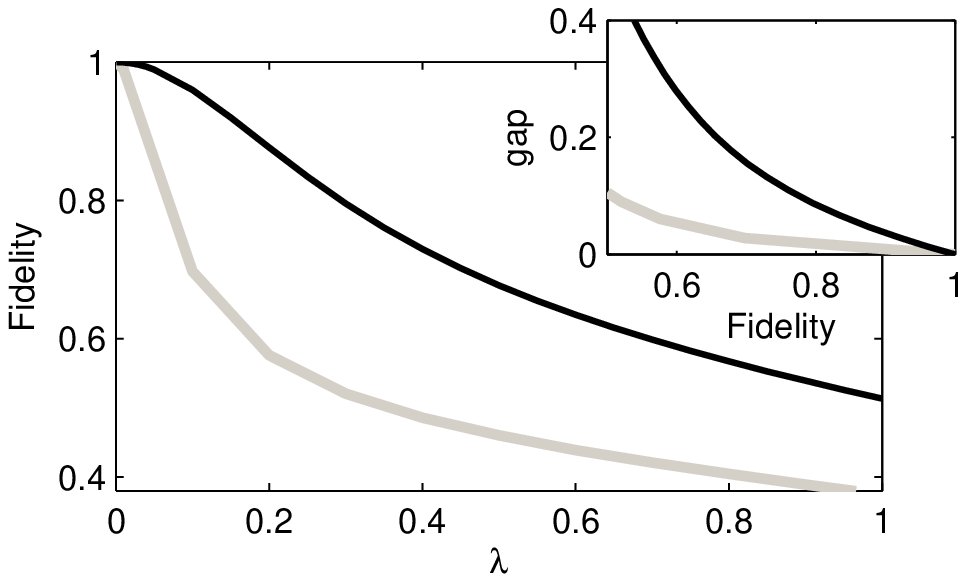}
\end{center}
\caption{Top: quantum spin networks with square (left) and cubic (right) configurations, each with four surface spins and with ferromagnetic interactions of the $XX$ type
($J_{jk}^x=J_{jk}^y=K^x_\ell=K^y_\ell=J$, $\forall j,k\in B$ and $\forall \ell\in S$). Bottom: fidelity between the multipartite entangled state in Eq.~\rp{psi0} and the reduced ground-state density matrix of the surface spins as a function of the surface-to-bulk coupling strength $\lambda$. Inset: behavior of the
corresponding energy gap as a function of $\lambda$.}
\label{fig8}
\end{figure}

In order to gain insight into the structure of multipartite entanglement in the reduced
ground-state density matrix of the surface spins (which is a multipartite mixed state), we have to
bypass the problem of the lack of well defined and faithful measures of multipartite entanglement
in multipartite mixed states. We will proceed first by introducing a reasonable ansatz (see Eq. (\ref{psi0}) for the pure ground state of the network's surface in the limit of small/vanishing coupling to the bulk, and we will discuss the structure of the associated residual tangle, which is a well defined and faithful measure of multipartite entanglement for pure states~\cite{Coffman}. We will then show how the reduced ground state of the network's surface indeed approaches the proposed ansatz state as the coupling $\lambda$ of the surface to the bulk is progressively reduced. If for instance we consider a network with $XX$ ferromagnetic interactions with a surface composed by four spins as illustrated in
Fig.~\ref{fig8}, then, following the discussion in Sec.~\ref{Model}, the effective surface dynamics is described by a fully connected effective $XX$ Hamiltonian for the four surface spins; therefore the natural pure multipartite ansatz for the network's surface is given by the ground state of the effective, fully connected surface Hamiltonian. For symmetric configurations, such as those illustrated in Fig.~\ref{fig8}, each spin of the surface will effectively interact with all the other surface spins, with the same coupling strength. It is straightforward to verify that the ground state of the corresponding effective Hamiltonian, in the ferromagnetic case, is the following multipartite entangled state:
\begin{eqnarray}\label{psi0}
\ke{Z_0}&=&\frac{1}{\sqrt{6}}\lpt{\ke{\up\up\dow\dow}+\ke{\up\dow\up\dow}+\ke{\up\dow\dow\up}
}\nn\\&&\rpt{
+\ke{\dow\up\up\dow}+\ke{\dow\up\dow\up}+\ke{\dow\dow\up\up}} \; .
\end{eqnarray}
By symmetry considerations, the ground state of the effective $4$-spin $XX$ Hamiltonian must have net zero magnetization on the $XY$ plane, and state $\ke{Z_0}$, Eq. (\ref{psi0}), is the unique state of lowest energy with this property, in accordance with the fact that for gapless systems the ground state is non degenerate. The normalized residual tangle, taking values in the interval $[0,1]$~\cite{Coffman} can be computed exactly for the state $\ke{Z_0}$, Eq. (\ref{psi0}), yielding $2/3$ for each spin~\footnote{The residual tangle, ${\cal R}_j$, for a spin $j$ is defined as the difference between the corresponding tangle,  ${\cal T}_j=4{\rm Det}\pq{\rho_1}$ where $\rho_j$ is the reduced density matrix of the spin $j$, and the sum of the square of the concurrences, ${\cal C}_{j,k}$ for $k\neq j$, between spin $j$ and all the other spins: ${\cal R}_j={\cal T}_j-\sum_{k\neq j}{\cal C}_{j,k}^2$.},
a large finite value that demonstrates the existence of strong multipartite entanglement between the surface spins. In the limit of very small values of the surface-to-bulk coupling $\lambda$, state $\ke{Z_0}$, Eq. (\ref{psi0}), approximates closely the reduced ground sate $\rho_0$ of the surface spins. This is shown in Fig.~\ref{fig8} where we report the fidelity $F=\Tr\pg{\rho_0\ke{Z_0}\br{Z_0}}$ as a function of the parameter $\lambda$. This behavior demonstrates the presence of strong genuine multipartite entanglement among the surface spins in networks with highly symmetric interactions.

\section{conclusion and outlook}
\label{Conclusions}

In the present work we have introduced and discussed the properties of surface entanglement, a generalization of the phenomenon of long-distance entanglement in quantum spin chains~\cite{Campos Venuti1, Campos Venuti2,GiampaoloLong1,GiampaoloLong2} to higher dimensional arrays of quantum spins.
Surface entanglement is defined as the entanglement present in the reduced ground state of distant and non directly interacting spins belonging to the external boundary of a quantum spin network. The conditions for the occurrence of sizeable surface entanglement are that the surface spins be weakly coupled to the bulk spins and that the ground state of the bulk be non degenerate.
We have observed that, typically, geometrically frustrated networks exhibit a weaker surface entanglement compared to the corresponding geometrically unfrustrated ones. We have also discussed how the surface entanglement is enhanced in networks with a modular structure, thereby extending the concept of one-dimensional modular entanglement~\cite{Gualdi} to structures with modules of arbitrary geometry and dimensionality.

While long-distance entanglement allows to entangle only two distant spins at the ends of a quantum spin chain, surface entanglement may involve large numbers of spins resulting in a great variety of entanglement patterns. Indeed, we showed that it is possible to entangle in parallel several spins pairs at the surface of the network or to create strong multipartite entanglement between all the spins in the surface. These extended entanglement properties in the outer regions of a network with weak boundary-to-bulk links are promising resources for routing quantum information among many distant nodes.
Indeed, the ground-state surface entanglement that we have demonstrated is the result of effective interactions between the surface spins that are mediated by and have the same symmetries of the strongly correlated bulk. In this context, an interesting and relevant question is whether it may be possible to control the effective surface dynamics for quantum information and communication purposes, in particular for the efficient, parallel state transfer between many nodes in a  network~\cite{Christandl2004,Friesen,Pemberton-Ross2011,Kay}.


\acknowledgments
We acknowledge financial support through the FP7 STREP Project iQIT, Grant Agreement n. 270843.

\end{document}